\documentclass[aps,pre,onecoloumn,showpacs]{revtex4-1}  
\usepackage[dvipsnames]{xcolor}
\usepackage{graphicx}  
\usepackage{dcolumn}   
\usepackage{bm}       
\usepackage{amssymb}  
\usepackage{amsmath}
\usepackage{dsfont}
\usepackage{float}
\usepackage{xcolor}
\usepackage{enumerate}
\usepackage{braket}
\begin{document}
\title{Facets of nonlocal correlation under non-Hermitian system}
\author{ J. Ramya Parkavi$^1$, R. Muthuganesan$^1$, and V. K. Chandrasekar$^1$}
\address{Department of Physics, Centre for Nonlinear Science \& Engineering, School of Electrical \& Electronics Engineering, SASTRA Deemed University, Thanjavur -613 401, Tamilnadu, India.}
 \begin{abstract} 
\par In this article, we investigate the  dynamics of a bipartite system under the action of a local non-Hermitian system. We study the quantum correlation of the bipartite system quantified by the entanglement, measurement-induced nonlocality (MIN) based on Hilbert$-$Schmidt norm, trace distance, and Bell inequality. We find that the quantum correlations of the system depend on the initial conditions and system parameters. We observe that the states with nonzero quantum correlation obey the Bell inequality even in the absence of entanglement.  Moreover, the Bell inequality completely fails to manifest the nonlocality for the mixed quantum state. However, we have identified the nonlocal attributes of the mixed quantum state in terms of MIN and trace distance MIN.   Our results show that the trace distance-based correlation is more robust against the nonunitary evolution compared to the other quantifiers.
\end{abstract} 
\maketitle
\section{Introduction}
Since the inception of quantum theory, it is believed that physical observables are Hermitian and having real eigenvalues. After the seminal work of Bender  \cite{nonhermpt1,nonhermpt2,nonhermmpt3}, the researchers understood that there are few non-hermitian systems also possessing real eigenvalues and generate nonunitary time evolution. Due to rapid development in both theoretical and experimental facts, the class of general non-Hermitian systems acts as an effective tool in the domain of quantum information theory to describe the dynamics of the quantum system which interacts with the environment \cite{nonhemenvir1,nonhemenvir2}. For instance, non-Hermitian systems have been used to explain the well-known dissipation and absorption phenomena like radioactive decay \cite{radiactive} and interaction between neutrons and nuclei \cite{neutron}. Characterizing the nonlocal attributes of a quantum system is a formidable task and a key to implement the various quantum information processing \cite{Nielsen2010}. To manifest the nonlocality, various measures with different notions have been introduced. A few popular measures are coherence \cite{Baumgratz}, entanglement \cite{Einstein,Schrodinger}, Bell inequality \cite{Bell}, steering \cite{Schneeloch}, etc,. In general, nonlocality of a state witnessed or demonstrated through the violation of Bell inequality. For any pure state, the entanglement and Bell inequality are the complete manifestation of nonlocality. On the other hand, the nonlocality of the mixed state is not entirely understood. Indeed, the entanglement and Bell inequality are the incomplete manifestation of nonlocality.  Due to this incompleteness, we are in need of identifying new quantum correlation measures, beyond entanglement.

To characterize the complete nonlocal attributes of the quantum system, in the past two decades, various quantum correlation measures have been identified and explored in detail \cite{Ollivier,Dakic,Luo2011,Luo2008,Luo2010,Xiong2014}. In geometric perspectives, based on the locally invariant projective measurement, a new measure of quantum correlation has been introduced which is known as measurement-induced nonlocality (MIN) \cite{Luo2011}. Originally, MIN is defined via maximal Hilbert-Schmidt norm, and considered as a more general manifestation of nonlocality than the Bell inequality and entanglement. The dual of this  quantity is also defined using Hilbert-Schmidt norm \cite{Dakic}.   Even though such measures were realizable in terms of local observable  \cite{Jin2012,Passante,Girolami2012}, Piani indicated that the bipartite correlation measures based on the Hilbert-Schmidt distance are not a faithful measure of quantum correlation \cite{Piani2012}. In order to circumvent the local ancilla issue, MIN has redefined using different distance measures \cite{Muthu1,Fan2015,Muthu2,indra2021} and entropic quantities \cite{Xi2012,Hu2012,Li2016}. In recent times, these quantities were studied in various physical systems such as Heisenberg spins \cite{indra1}, optical systems \cite{Mohamed1}, and diamond spin chain \cite{Bhuvana}. 

In general, the interaction between a quantum system and the environment is unavoidable and the system loses its unique quantum mechanical property. For an efficient information processing application, the quantum resources like quantum correlation should be protected against the environmental effect or decoherence. As a matter of fact, the dynamics of quantum correlation in Hermitian systems under decoherence and intrinsic decoherence is quite well studied \cite{Mohamed1,Mohamed2,Muthu2021QIP}. The dynamical behaviors of quantum correlation under the non-Hermitian system are still not addressed. 

To address the above, in this article, we study the nonunitary evolution of quantum correlations quantified by the entanglement, maximal Bell function and MIN. It is evident that the Bell inequality and entanglement are incomplete manifestations of nonlocality for the non-Hermitian system also. The nonunitary evolution causes sudden death in entanglement and Bell function, whereas the MINs are more robust against decoherence and exhibits the frozen dynamics. 

The article is prepared in the following manner: First, we provide the overview of different measures of quantum correlation in Sec. \ref{sec2}. Then, we present the considered system in Sec. \ref{sec3}. Further, we study the quantum correlation measures for the pure state in Sec. \ref{sec4} A. Then, we discuss the correlation measures of the mixed state in Sec. \ref{sec4} B. Finally, we conclude the results in Sec.\ref{cncl}.
 
\section{Nonlocal correlation quantifiers}
\label{sec2}
In this section, we review the quantum correlation measures examined in this article. 

\textit{Entanglement:}
Let $\rho$ be a bipartite quantum state shared by the marginal states $\rho^a$ and $\rho^b$. The degree of entanglement between the subsystems is quantified by the concurrence and is defined as \cite{Wootters} 
\begin{align}
\mathcal{C}(\rho)=2~ \text{max}~\left\{0,~ 2\epsilon_1-\sum_{i=1}^4 \epsilon_i \right\},
\end{align}
where $\epsilon_i$ are the square root of eigenvalues of matrix $R=\sqrt{\rho}\tilde{\rho} \sqrt{\rho}$ and arranged in decreasing order i.e., $\epsilon_1\geq \epsilon_2\geq \epsilon_3 \geq \epsilon_4$. Here $\tilde{\rho}=(\sigma_y \otimes \sigma_y)\rho^*(\sigma_y \otimes \sigma_y)$ is the spin flipped density operator and the symbol $*$ denotes the usual complex conjugate in the computational basis. The well-known fact is that the concurrence varies from $0$ to $1$ with minimum and maximum values correspond to unentangled and maximally entangled states, respectively.

\textit{Measurement-induced nonlocality:}
The Bloch representation of the bipartite density operator is  introduced by Fano. The general expression of a two-qubit density operator acting in the separable composite Hilbert space
$\mathcal{H}^a\otimes\mathcal{H}^b$ is 
\begin{align}
\rho=\frac{1}{4}\left(\mathds{1}^a \otimes \mathds{1}^b+ \mathbf{x} \cdot  \boldsymbol\sigma \otimes \mathds{1}^b + \mathds{1}^a \otimes \mathbf{y}\cdot\boldsymbol\sigma+\sum_{m,n= 1}^3 t_{mn} \sigma_m \otimes \sigma_n\right), \label{Equations} 
\end{align}
where $\sigma_j ~(j=1,2,3)$ are the Pauli operators. The vectors $\mathbf{x}$ and $\mathbf{y}$ are real, their components being $x_j=\text{Tr}(\rho (\sigma_j\otimes \mathds{1}))$, $\mathds{1}^{a(b)}$ is the $2\times 2$ unit operator acting on the subsystem $a(b)$ and $y_j=\text{Tr}(\rho (\mathds{1}\otimes \sigma_j ))$, $t_{mn}=\text{Tr}(\rho(\sigma_m \otimes \sigma_n))$ being real matrix elements of correlation matrix $T$ and $\text{Tr}$ denote trace of a matrix. Without loss of generality, the canonical form of the Fano parametrization of the density operator (\ref{Equations})  is written as 
\begin{align}
\rho=\frac{1}{4}\left(\mathds{1}^a \otimes \mathds{1}^b+ \mathbf{x} \cdot  \boldsymbol\sigma \otimes \mathds{1}^b + \mathds{1}^a \otimes \mathbf{y}\cdot\boldsymbol\sigma+\sum_{j= 1}^3 c_{j} \sigma_j \otimes \sigma_j\right) \label{state2} 
\end{align}
with $c_j=\text{Tr}(\rho(\sigma_j \otimes \sigma_j))$.

MIN, captures the maximal nonlocal effects of bipartite state due to locally invariant projective measurements,  is defined as the maximal distance between the quantum state of our consideration and the corresponding state after performing a local measurement on one of the subsystems, say $a$ i.e.,\cite{Luo2011}
\begin{equation}
 \mathcal{N}_{2}\mathcal{(\rho)}=~^{\text{max}}_{\Pi ^{a}}\| \rho - \Pi ^{a}(\rho )\| ^{2},
\end{equation}
where $\|\mathcal{O}\| ^{2}=\text{Tr}(\mathcal{O}\mathcal{O}^{\dagger})$ is Hilbert-Schmidt norm of operator $\mathcal{O}$ and the maximum is taken over the locally invariant projective measurements on subsystem $a$ which does not change the state.  The post-measurement is defined as $\Pi^{a}(\rho) = \sum _{k} (\Pi ^{a}_{k} \otimes   \mathds{1} ^{b}) \rho (\Pi ^{a}_{k} \otimes    \mathds{1}^{b} )$, with $\Pi ^{a}= \{\Pi ^{a}_{k}\}= \{|k\rangle \langle k|\}$ being the projective measurements on the subsystem $a$, which do not change the marginal state $\rho^{a}$ locally i.e., $\Pi ^{a}(\rho^{a})=\rho ^{a}$. If $\rho^{a}$ is non-degenerate, then the maximization is not required. For a given state Eq. (\ref{Equations}), the MIN has a closed formula as 
\begin{equation}
\mathcal{N}_{2}\mathcal{(\rho)} =
\begin{cases}
\text{Tr}(TT^t)-\frac{1}{\| \textbf{x}\| ^2}\textbf{x}^tTT^t\textbf{x}& 
 \text{if} \quad \textbf{x}\neq 0,\\
 \text{Tr}(TT^t)- T_3&  \text{if} \quad \textbf{x}=0,
\end{cases}
\label{HSMIN}
\end{equation}
where $T_3$ is the least eigenvalue of matrix $TT^t$, the superscripts $t$ stands for the transpose of a matrix.

\textit{Trace distance-based MIN}:
It is a well-known fact that the MIN based on the Hilbert-Schmidt norm is not a bonafide measure in capturing nonlocal attributes of a quantum state due to the local ancilla problem \cite{Piani2012}. A  natural way to circumvent this issue is defining MIN in terms of contractive distance measure. Another alternate form of MIN is based on trace distance \cite{Fan2015},  namely, trace MIN (T-MIN)  which resolves the local ancilla problem  \cite{Piani2012}. 
It is defined as  \cite{Fan2015}
\begin{equation}
\mathcal{N}_{1}\mathcal{(\rho)}:= ~^{\text{max}}_{\Pi^a}\Vert\rho-\Pi^a(\rho)\Vert_1,
\end{equation} 
where $\Vert \mathcal{O} \Vert_1 = \text{Tr}\sqrt{\mathcal{O}^{\dagger}\mathcal{O}}$ is the trace norm of operator $\mathcal{O}$. Here also, the maximum is taken over all locally invariant projective measurements. For the considered system, the closed formula of trace MIN $\mathcal{N}_{1}\mathcal{(\rho)}$ is given as 
\begin{equation}
\mathcal{N}_{1}\mathcal{(\rho)}=
\begin{cases}
\frac{\sqrt{\chi_+}~+~\sqrt{\chi_-}}{2 \Vert \textbf{x} \Vert_1} & 
 \text{if} \quad \textbf{x}\neq 0,\\
\text{max} \lbrace \vert c_1\vert,\vert c_2\vert,\vert c_3\vert\rbrace &  \text{if} \quad \textbf{x}=0,
\end{cases}
\label{TMIN}
\end{equation}
where $\chi_\pm~=~ \alpha \pm 2 \sqrt{\tilde{\beta}}~\Vert \textbf{x} \Vert_1 ,\alpha =\Vert \textbf{c} \Vert^2_1 ~\Vert \textbf{x} \Vert^2_1-\sum_i c^2_i x^2_i,\tilde{\beta}=\sum_{\langle ijk \rangle} x^2_ic^2_jc^2_k,~ \vert c_i \vert $ are the absolute values of $c_i$ and the summation runs over cyclic permutation of 
$\lbrace 1,2,3 \rbrace$.

\textit{Maximal Bell function:}
  To manifest another form of  bipartite nonlocality, we employ the Bell maximal function $\mathcal{B}(\rho)$ as a another measure of nonclassical correlation. In 1964, J. Bell demonstrated that violation of the Bell inequality by a bipartite state using a local hidden variable \cite{Bell}. In general, the violation of Bell inequality witnesses the nonlocal attributes of the quantum system. The Bell function is defined as \cite{Brunner,Horodecki}
\begin{align}
  \mathcal{B}(\rho)=\sqrt{2} \text{max}[T_{1}+T_{2}, T_{2}+T_{3}, T_{1}+T_{3}],
\end{align}
where  $T_{i}(i=1,2,3)$ are the eigenvalues of the matrix $T^{t}T$. The function $\mathcal{B}(\rho)>2$ quantifies  the  amount  of  Bell  inequality  violation  and the nonlocal correlation of two-qubit quantum states.

\section{Model and Evolution}
\label{sec3}
In order to understand the dynamical properties of different quantum correlation measures under non-unitary evolution, we consider the Hamiltonian of a single qubit non-Hermitian system, it can be written as \cite{baseref}
\begin{equation}
 H=-i\gamma \sigma_{z}-\Delta \sigma_{x},
 \label{ham}
\end{equation}
where  $\Delta$ and $\gamma$ are the real parameters, $\sigma_{x}$ and $\sigma_{z}$ are usual Pauli spin operators. The first term in the above equation represents the decay rate operator and the second term implies the qubit which undergoes coherent Rabi oscillation. Here, $\Delta$ represents the coupling strength, and $\gamma$ characterize the non-Hermiticity of the Hamiltonian. The eigenvalues of the Hamiltonian given in Eq. (\ref{ham}) are $E_{1,2}={\pm}\sqrt{\Delta^{2}-\gamma^{2}}$. The eigenvalues are real when $\Delta<\gamma$ and turn to be imaginary when $\gamma$ value exceeds $\Delta$.   
 The time evolution of arbitrary quantum state $\rho(0)$ can also be represented as
 \begin{equation}
 \rho(t)=U(t)\rho(0)U^{\dagger}(t) \nonumber
 \end{equation}
 by introducing a time-evolution operator
 \begin{equation}
 U(t)=e^{-iHt}. \nonumber
 \end{equation}
 Note that $\rho(t)$ is not a trace preserving here due to the non-Hermiticity of Hamiltonian, while observers have to measure the quantum system in a conventional quantum world. Thus, $\rho(t)$ has to be renormalized:
 \begin{equation}
 \tilde\rho(t)=\frac{\rho(t)}{Tr (\rho(t))} \nonumber.
 \end{equation}
To characterize the bipartite correlation, we consider the two-qubit initial state $\rho(0)$, Then the  evolved state will be:
\begin{equation}
\tilde{\rho}(t)=\frac{(U^{a}(t) \otimes \mathds{1}^b)\rho(0)(U^{a}(t) \otimes \mathds{1}^b)^\dagger}{M}, 
\label{pt}
\end{equation}
where $M=Tr(\rho(t))$ is the normalization parameter. For any two-qubit initial state, we obtain the time evolved state as
\begin{equation}
\tilde{\rho}(t)=\frac{1}{M}
\begin{pmatrix}
\rho_{11} & \rho_{12} & \rho_{13} & \rho_{14} \\
\rho_{21} & \rho_{22} & \rho_{23} & \rho_{24} \\
\rho_{31} & \rho_{32} & \rho_{33} & \rho_{34} \\
\rho_{41} & \rho_{42} & \rho_{43} & \rho_{44} \\
\end{pmatrix}.
\label{mainrho}
\end{equation}

\begin{figure*}[t]
	\centering
	\includegraphics[width=1.0\textwidth]{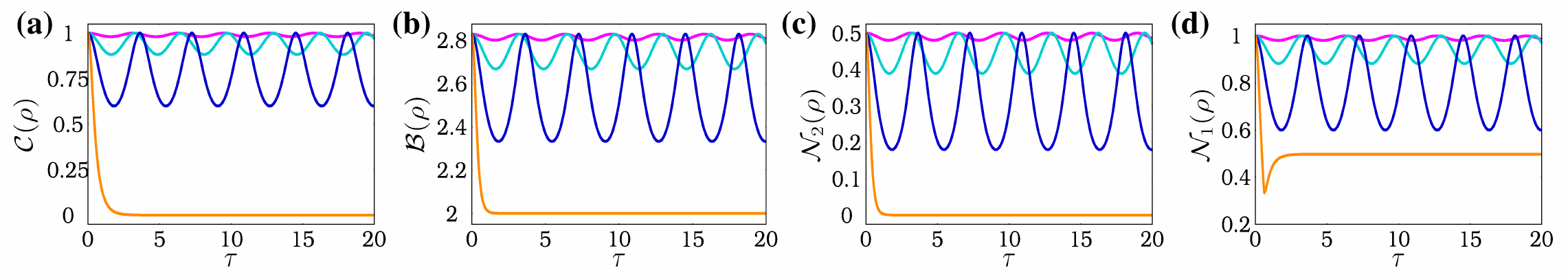}
	\caption{Time evolution of quantum correlations for the pure state $\ket{\psi (0)}=\frac{1}{\sqrt{2}}(\ket{00}+\ket{11})$ under non-Hermitian system for different values of $\tilde{\gamma}$: $\tilde{\gamma}=0.1$ (magenta solid line), $\tilde{\gamma}=0.25$ (lightblue solid line), $\tilde{\gamma}=0.5$ (dark blue solid line), and $\tilde{\gamma}=1.5$ (orange solid line). (a) shows Concurrence, (b) shows Bell inequalities, (c) shows MIN, and (d) shows Trace MIN.}
	\label{aeqlb0011}
\end{figure*}
\begin{figure*}[t]
	\centering
	\includegraphics[width=1.0\textwidth]{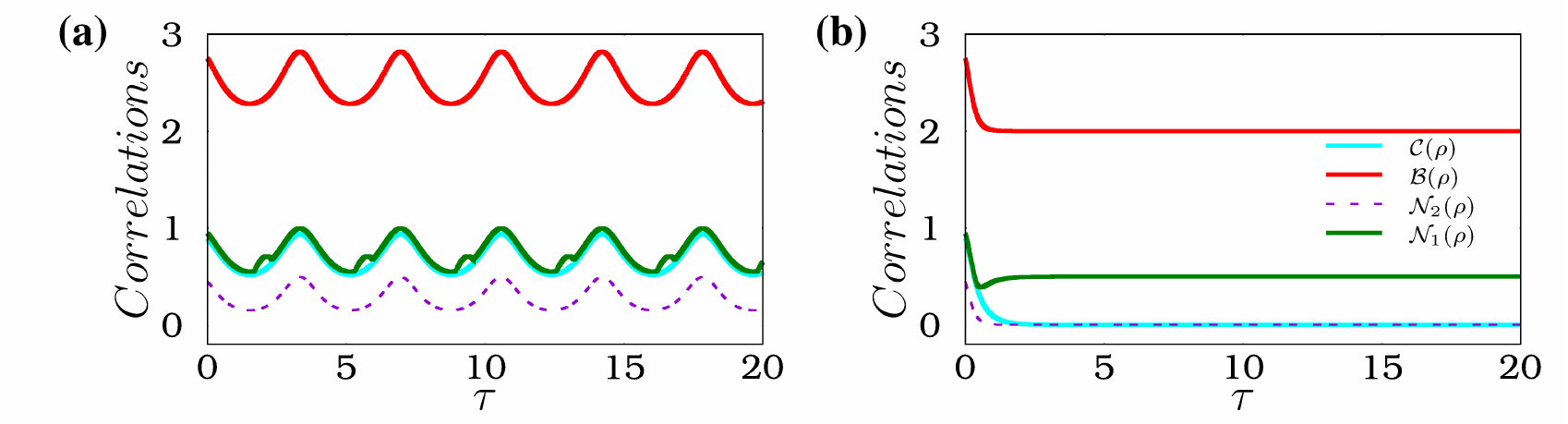}
	\caption{Fig. (a) and (b) combinely shows the MIN (purple dashed line), Trace MIN (darkgreen solid line), Bell inequalities (Red solid line) and Concurrence (cyan solid line) for two different values of $\tilde{\gamma}=0.5 \text{and} 1.5$ for an initally pure state $\ket{\psi (0)}=\frac{1}{\sqrt{3}}\ket{00}+\sqrt{\frac{2}{3}}\ket{11}$.}
	\label{anoteqlb0011}
\end{figure*}
\section{Results and discussions}
\label{sec4}
\subsection{Pure state}

In what follows, we focus on the dynamics of bipartite quantum correlation quantified by the entanglement, measurement-induced nonlocality (MIN), and Bell nonlocality. For this purpose, we consider a pair of initially correlated qubits as initial state $\rho(0)=a\ket{00}+b\ket{11}$ with  $a$ and $b$ are the complex coefficients satisfies the normalization condition $|a|^{2}+|b|^{2}=1$. The matrix elements of the time evolved state are
\begin{eqnarray}
 && \rho_{11}=a^2 \left(\cosh \mu  \tau-\frac{\tilde{\gamma} \sinh \mu \tau}{\mu }\right)^2,  ~~
  \rho_{22}=\frac{b^2 \sinh ^{2} \mu  \tau }{\mu ^2},
 \rho_{33}=\frac{a^2 \sinh ^{2}\mu  \tau }{\mu ^2},
  \rho_{44}=b^2 \left(\cosh \mu  \tau +\frac{\tilde{\gamma} \sinh \mu \tau}{\mu } \right)^2, \nonumber\\
 &&\rho_{12}=\rho_{21}^*= \frac{i a b \sinh \mu  \tau (\tilde{\gamma} \sinh \mu \tau-\mu  \cosh \mu  \tau )}{\mu ^2}, 
 \rho_{13}=\rho_{31}^*=\frac{i a^2 \sinh \mu \tau (\tilde{\gamma} \sinh \mu \tau-\mu  \cosh \mu  \tau )}{\mu ^2}, \nonumber \\
&&\rho_{14}=\rho_{41}=a b \left(\cosh ^{2} \mu  \tau-\frac{\tilde{\gamma}^2 \sinh ^{2} \mu  \tau }{\mu ^2}\right), \rho_{24}=\rho_{42}^*=\frac{i b^2 \sinh \mu \tau (\tilde{\gamma} \sinh \mu \tau+\mu  \cosh \mu  \tau )}{\mu ^2}, \nonumber\\
 &&\rho_{23}=\rho_{32}=\frac{a b \sinh ^{2} \mu  \tau }{\mu ^2}, 
\rho_{34}=\rho_{43}^*=\frac{i a b \sinh \mu \tau (\tilde{\gamma} \sinh \mu \tau+\mu  \cosh \mu  \tau )}{\mu ^2}, 
\end{eqnarray}
where, $\tau=\Delta t$, $\tilde{\gamma}=\frac{\gamma}{\Delta}$, and $\mu=\sqrt{\tilde{\gamma}^{2}-1}$. The normalization parameter is $
M=M_{1}+M_{2}$
with, 
\begin{align}
M_1=a^2 \left(\cosh \mu  \tau -\frac{\tilde{\gamma} \sinh \mu \tau}{\mu }\right)^2+\frac{b^2 \sinh ^{2}\mu  \tau }{\mu ^2},\nonumber  ~ 
M_2=\frac{a^2 \sinh ^{2}\mu  \tau }{\mu ^2}+b^2 \left(\cosh \mu  \tau +\frac{\tilde{\gamma} \sinh \mu \tau}{\mu }\right)^2.	 \nonumber
\end{align}

To analyze the evolution of quantum correlations under nonunitary evolution, we set $a=b=1/\sqrt{2}$ and the initial state is a maximally entangled state. Fig. \ref{aeqlb0011} depicts the dynamical behaviors of quantum correlation measures for the exemplary values of $\tilde{\gamma}~(= 0.1,~0.25,~0.5,~ \text{and} ~1.5)$. At $t=0$, the concurrence $\mathcal{C}(\rho)$, $\mathcal{B}(\rho)$ and measurement-induced nonlocality (MIN) are maximum, it is obvious that the initial state is maximally entangled (correlated). Further, we observe that the trace MIN and concurrence are identical in capturing quantum correlation. A similar observation is made in Heisenberg spin (Hermitian) system under intrinsic decoherence as well \cite{Muthu2021QIP}.  For the lower values of $\tilde{\gamma}$, we observe that the MINs, concurrence, and Bell function oscillate with time.  At a given instant of time, the function $\mathcal{B}(\rho)>2$, implies the existence of nonlocal aspects in the considered quantum system. Further, we observe that the considered measures are similar in quantifying nonclassical correlation. 

The increment in $\tilde{\gamma}$ factor accelerates the two-qubit oscillating correlation to reach the steady state value rapidly compared to the lower values of $\tilde{\gamma}$. The measures $\mathcal{C}(\rho)$,  and $\mathcal{N}_{2}(\rho)$ are vanishing after a finite time, whereas $\mathcal{B}(\rho)$ saturates at 2 and this fact refers to the  sudden death of entanglement. Further, the trace MIN $N_1(\rho)$ saturates at 0.5, implying that the trace MIN is more pronounced than the other companion quantities.  In general, the entangled state witnessed through the violation of Bell inequality. From Fig. \ref{aeqlb0011}(b), we observe that after some finite time the Bell function is always equal to 2 which indicates that the quantum state has no nonclassical correlation and the corresponding quantum state obeys the Bell inequality. This observation is also reflected in the dynamics of entanglement. Whereas the nonzero trace MIN indicates the presence of quantum correlation in the separable quantum state.  Hence, trace MIN is a good measure of quantumness compared to entanglement measure.
\begin{figure*}[t]
	\centering
	\includegraphics[width=1.0\textwidth]{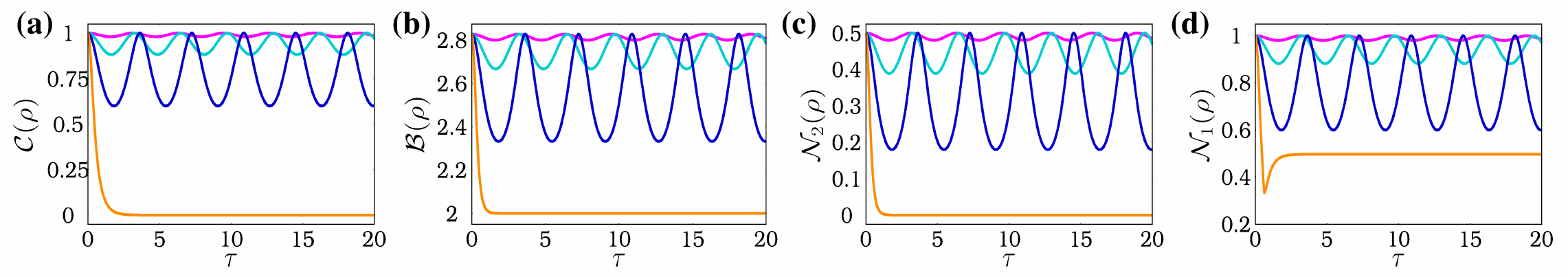}
		\caption{Time evolution of quantum correlations for the pure state $\ket{\psi (0)}=\frac{1}{\sqrt{2}}(\ket{01}+\ket{10})$ under non-Hermitian system for different values of $\tilde{\gamma}$:  $\tilde{\gamma}=0.1$ (magenta solid line), $\tilde{\gamma}=0.25$ (lightblue solid line), $\tilde{\gamma}=0.5$ (dark blue solid line), and $\tilde{\gamma}=1.5$ (orange solid line). (a) shows Concurrence, (b) shows Bell inequalities, (c) shows MIN, and (d) shows Trace MIN.}
	\label{aeqlb0110}
\end{figure*}

Next, we consider the above pure state with another initial condition $a=\frac{1}{\sqrt{3}}$ and $b=\sqrt{\frac{2}{3}}$. Substituting the values of $a$ and $b$ in Eq. (\ref{mainrho}), one can obtain the matrix elements of time evolved state and characterize the dynamical aspects of quantum correlation. We have plotted the correlation measures in Fig. \ref{anoteqlb0011} for the fixed values of $\tilde{\gamma}$. Here also, we observe that all the correlation measures oscillating with time for the value of  $\tilde{\gamma}=0.5$ as shown in Fig. \ref{anoteqlb0011}(a). For higher values of $\tilde{\gamma}$ say  $\tilde{\gamma}=1.5$, the functions  $\mathcal{N}_{2}\mathcal{(\rho)}$ and $\mathcal{C}(\rho)$ decreases monotonically, then reaches to zero. As shown in the red curves of Fig. 2(b),  the maximal Bell function $\mathcal{B}(\rho)$ saturates at 2 after a finite time and fails to manifest the nonlocality. Further, we observe that the rapid decaying behavior of $\mathcal{N}_{1}\mathcal{(\rho)}$  and the frozen dynamics after a shorter time. The above observation emphasizes that the absence of entanglement (obey the Bell inequality)  not necessarily indicates the absence of nonlocality of a quantum system. 

To assert the pure state dynamics of quantum correlation, next, we consider another form of pure state $\ket{\psi}(0)=a\ket{01}+b\ket{10}$,  $a$ and $b$ are complex numbers and $a^{2}+b^{2}=1$. The corresponding density matrix elements are

\begin{eqnarray}
&& \rho_{11}=\frac{ b^{2}\sinh ^2(\mu  \tau )}{\mu ^2}, \rho_{22}=a^{2}\left(\cosh \mu  \tau -\frac{\tilde{\gamma} \sinh \mu \tau}{\mu }\right)^2, 
 \rho_{33}=b^{2}\left(\cosh \mu  \tau +\frac{\tilde{\gamma} \sinh \mu \tau}{\mu }\right)^2, \rho_{44}=\frac{ a^{2}\sinh ^2(\mu  \tau )}{\mu ^2},\nonumber\\
 && \rho_{12}=\rho_{21}^*= \frac{i a b \sinh \mu  \tau (\tilde{\gamma} \sinh \mu \tau-\mu  \cosh \mu  \tau )}{\mu ^2}, 
 \rho_{13}=\rho_{31}^*=\frac{i b^2 \sinh \mu \tau (\tilde{\gamma} \sinh \mu \tau+\mu  \cosh \mu  \tau )}{\mu ^2}, \nonumber \\
&&
 \rho_{23}=\rho_{32}=a b \left(\cosh ^{2} \mu  \tau-\frac{\tilde{\gamma}^2 \sinh ^{2} \mu  \tau }{\mu ^2}\right), \rho_{24}=\rho_{42}^*=\frac{i a^2 \sinh \mu \tau (\tilde{\gamma} \sinh \mu \tau-\mu  \cosh \mu  \tau )}{\mu ^2}, \nonumber \\
&&  \rho_{14}=\rho_{41}=\frac{a b \sinh ^{2} \mu  \tau }{\mu ^2}, ~
 \rho_{34}=\rho_{43}^*=-\frac{i a b \sinh \mu \tau (\tilde{\gamma} \sinh \mu \tau+\mu  \cosh \mu  \tau )}{\mu ^2}. 
\end{eqnarray}
\begin{figure*}[t]
	\centering
	\includegraphics[width=1.0\textwidth]{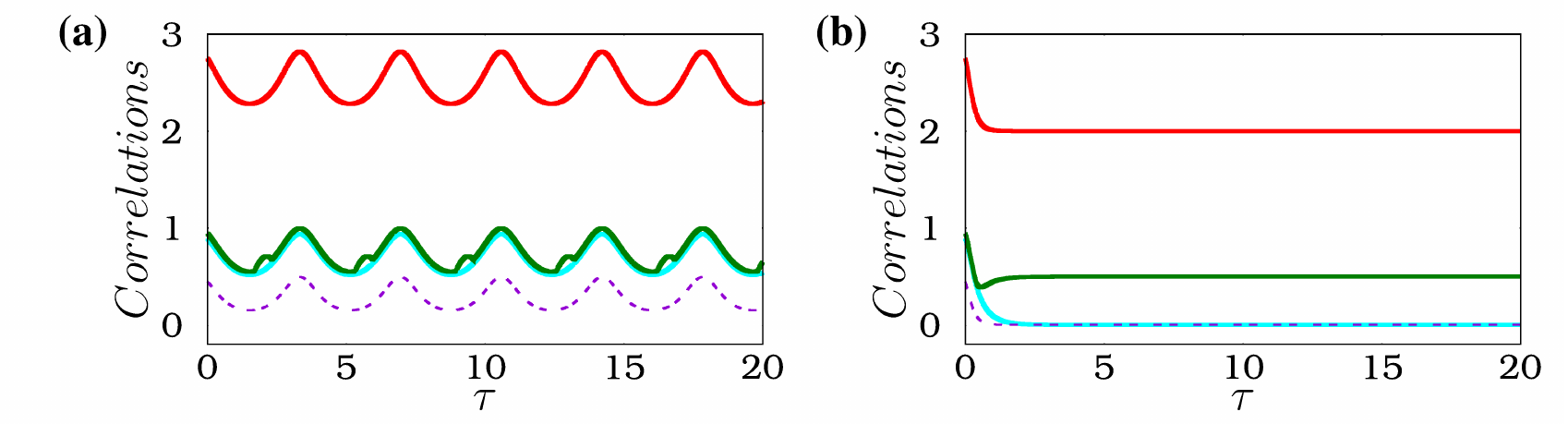}
	\caption{Fig. (a) and (b) combinely shows the MIN (purple dashed line), Trace MIN (darkgreen solid line), Bell inequalities (Red solid line) and Concurrence (cyan solid line) for two different values of $\tilde{\gamma}=0.5,1.5$ for an initally pure state $\ket{\psi (0)}=\frac{1}{\sqrt{3}}\ket{01}+\frac{\sqrt{2}}{\sqrt{3}}\ket{10}$.}
	\label{anoteqlb0110}
\end{figure*}
At initial time $t=0$, all the correlation measures for the initial conditions $a=b=1/\sqrt{2}$ of the state $\ket{\psi}(0)=a\ket{01}+b\ket{10}$ are maximum, which are illustrated in Fig 3. As time increases, all the measures are exhibiting similar periodic patterns which are observed in the previous case and implying that the measures are identical in quantifying quantum correlations. Here also, we observe that the higher values of $\tilde{\gamma}$ accelerate the decay rate of quantum correlations from the maximal value to steady state. The quantum correlation is quantified by the concurrence and MIN vanishes after a finite time, whereas the trace MIN reaches a non-zero steady state value. This nonzero steady state correlation has practical implications in quantum information processing. Again, the Bell inequality saturates at 2 and reflects the same observation from that of entanglement.

To understand further, we consider the initial conditions $a=\frac{1}{\sqrt{3}}$ and $b=\sqrt{\frac{2}{3}}$. Also, in this case, the functions $\mathcal{C}(\rho)$ and $\mathcal{N}_{2}\mathcal{(\rho)}$ are oscillatory with nearly equal amplitudes in time for $\tilde{\gamma}=0.5$. For larger values of $\tilde{\gamma}$, the measures $\mathcal{C}(\rho)$ and $\mathcal{N}_{2}\mathcal{(\rho)}$ vanishes after a finite time. As shown in the red curves of Fig. \ref{anoteqlb0110}(a) and \ref{anoteqlb0110}(b), the values of $\mathcal{B}(\rho)$ is maximum at $t=0$, implying that the considered quantum states have a correlation between the marginal states.  But for larger values of $\tilde{\gamma}$ (say $\tilde{\gamma}=1.5$), the scenario is different where the red curve decays to a constant value as depicted in Fig. 4(b). In a similar manner, the Trace MIN $\mathcal{N}_{1}\mathcal{(\rho)}$ oscillates for lower values of $\tilde{\gamma}$ and reaches a steady state for higher values of $\tilde{\gamma}$.

\subsection{Mixed state}
As we mentioned earlier, the characterization of mixed state from the perspectives of quantum correlation is still not understood and entanglement is believed to be an incomplete manifestation of nonlocality. In order to study the complete nonlocal behaviors of mixed state, we consider an entangled mixed state 
\begin{align}
  \rho(0)=r\ket{\psi (0)}\bra{\psi (0)}+(\frac{1-r}{4})\mathds{1}_{4},
\end{align}
where  $r$ characterize the purity of the state, $\ket{\psi}(0)=\frac{1}{\sqrt{2}}(\ket{00}+\ket{11})$ and $\mathds{1}_{4}$ is a $4\times 4$ identity matrix. If $r=1$, the state is maximally entangled and $r=0$, the state is maximally mixed. The evolved density matrix  has the structure is given in Eq. (11). 
The diagonal elements are,\\
\begin{eqnarray}
&&  \rho_{11}=\frac{(1+r) \mu^{2} \cosh ^2(\mu \tau)+(1+\tilde{\gamma} ^2+r(\tilde{\gamma}^2-1))\sinh ^2(\mu \tau)-(1+r)\tilde{\gamma} \mu \sinh(2 \mu \tau)}{4 \mu ^2}, \nonumber \\
 && \rho_{22}=\frac{(1+r) \sinh ^2(\mu  \tau )+(1-r) (\mu  \cosh (\mu  \tau )-\tilde{\gamma} \sinh \mu \tau)^2}{4 \mu ^2}, \nonumber \\
&&  \rho_{33}=\frac{(1+r) \sinh ^2(\mu  \tau )+(1-r) (\mu \cosh (\mu  \tau )+\tilde{\gamma} \sinh \mu \tau)^2}{4 \mu ^2}, \nonumber \\
&&  \rho_{44}=\frac{(1+r) \mu^{2} \cosh ^2(\mu \tau)+(1+\tilde{\gamma} ^2+r(\tilde{\gamma}^2-1))\sinh ^2(\mu \tau)+(1+r)\tilde{\gamma} \mu \sinh(2 \mu \tau)}{4 \mu ^2},
\end{eqnarray}
and the off-diagonal elements are 
 \begin{eqnarray}
&& \rho_{12}=\rho_{21}^*=\frac{i r \sinh (\mu \tau) (\tilde{\gamma} \sinh (\mu \tau)-\mu  \cosh (\mu  \tau ))}{2 \mu ^2}, ~
 \rho_{13}=\rho_{31}^*=\frac{i \sinh \mu \tau (\tilde{\gamma} \sinh \mu \tau-\mu  r \cosh (\mu  \tau ))}{2 \mu ^2}, \nonumber \\
&& \rho_{23}=\rho_{32}=\frac{r \sinh ^2(\mu  \tau )}{2 \mu ^2}, 
 \rho_{24}=\rho_{42}^*=\frac{i \sinh \mu \tau (\tilde{\gamma} \sinh \mu \tau+\mu  r \cosh (\mu  \tau ))}{2 \mu ^2},  \nonumber \\
&&\rho_{34}=\rho_{43}^*=\frac{i r \sinh \mu \tau (\tilde{\gamma} \sinh \mu \tau+\mu  \cosh (\mu  \tau ))}{2 \mu ^2}.
 \end{eqnarray}

\begin{figure*}[t]
	\centering
	\includegraphics[width=1.0\textwidth]{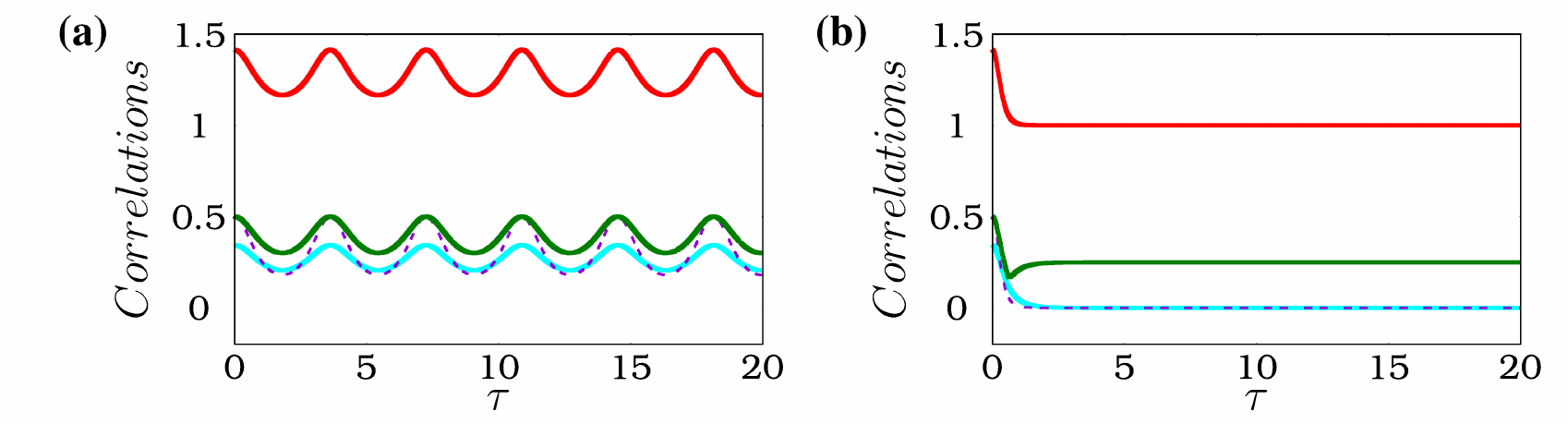}
	\caption{Fig. (a) and (b) shows the comparison of MIN (purple dashed line), Trace MIN (darkgreen solid line), Bell inequalities (red solid line) and Concurrence (cyan solid line) for two different values of $\tilde{\gamma}=0.5,1.5$ with $r=0.5$ for a mixed state $\ket{\psi (0)}=\frac{1}{\sqrt{2}}(\ket{00}+\ket{11})$.}
	\label{mixed0011}
\end{figure*}

\par To understand the role of factor $r$, we plot the MIN and Bell nonlocality of the considered system in Fig. 5 for fixed values of $\tilde{\gamma}=0.5,1.5$. We observe that the Bell function $\mathcal{B}(\rho)$ is always less than 2, implying that there is no nonclassical correlation between its local constituents. On the other hand, the companion quantities (Concurrence, MIN, and T-MIN) are nonzero while the state obeys the Bell inequality. In the realm of MIN, the separable (uncorrelated) state also possesses the nonlocal correlation and quantum advantages. This nonclassical correlation of local states is quantified by the MINs. This also reveals the fact that the Bell nonlocality is an incomplete manifestation of nonlocality. Here again, for higher values of $\tilde{\gamma}$, we observe similar decaying behavior and frozen dynamics of correlation measure $\mathcal{N}_{1}(\rho)$. However, $\mathcal{C}(\rho)$ and $\mathcal{N}_{2}(\rho)$ tends to be zero. One can observe the similar effects for lower values of $r$.

\par Next, we consider a  initial mixed state $r\ket{\psi (0)}\bra{\psi (0)}+(\frac{1-r}{4})I_{4}$. Here, if $\ket{\psi (0)}=\frac{1}{\sqrt{2}}(\ket{01}+\ket{10})$, then the matrix elements are given as 
\begin{eqnarray}
 && \rho_{11}=\frac{(1+r) \sinh ^2(\mu  \tau )+(1-r) (\mu  \cosh (\mu  \tau )-\tilde{\gamma} \sinh \mu \tau)^2}{4 \mu ^2}, \rho_{12}=\rho_{21}^*=\frac{i r \sinh (\mu \tau) (\tilde{\gamma} \sinh (\mu \tau)-\mu  \cosh (\mu  \tau ))}{2 \mu ^2}  \nonumber \\
&&  \rho_{22}=\frac{(1+r) \mu^{2} \cosh ^2(\mu \tau)+(1+\tilde{\gamma} ^2+r(\tilde{\gamma}^2-1))\sinh ^2(\mu \tau)-(1+r)\tilde{\gamma} \mu \sinh(2 \mu \tau)}{4 \mu ^2}, \rho_{23}=\rho_{32}=\frac{r \sinh ^2(\mu  \tau )}{2 \mu ^2}, \nonumber \\
 &&  \rho_{44}=\frac{(1+r) \sinh ^2(\mu  \tau )+(1-r) (\mu \cosh (\mu  \tau )+\tilde{\gamma} \sinh \mu \tau)^2}{4 \mu ^2},  \rho_{12}=\rho_{21}^*=\frac{i r \sinh (\mu \tau) (\tilde{\gamma} \sinh (\mu \tau)-\mu  \cosh (\mu  \tau ))}{2 \mu ^2},\nonumber \\
&& \rho_{13}=\rho_{31}^*=\frac{i \sinh \mu \tau (\tilde{\gamma} \sinh \mu \tau+\mu  r \cosh (\mu  \tau ))}{2 \mu ^2} , \rho_{24}=\rho_{42}^*=\frac{i \sinh \mu \tau (\tilde{\gamma} \sinh \mu \tau-\mu  r \cosh (\mu  \tau ))}{2 \mu ^2} \nonumber \\
&& \rho_{33}=\frac{(1+r) \mu^{2} \cosh ^2(\mu \tau)+(1+\tilde{\gamma} ^2+r(\tilde{\gamma}^2-1))\sinh ^2(\mu \tau)+(1+r)\tilde{\gamma} \mu \sinh(2 \mu \tau)}{4 \mu ^2}\nonumber \\
&& \rho_{34}=\rho_{43}^*=-\frac{i r \sinh \mu \tau (\tilde{\gamma} \sinh \mu \tau+\mu  \cosh (\mu  \tau ))}{2 \mu ^2}.
\end{eqnarray}

\begin{figure*}[t]
	\centering
	\includegraphics[width=1.0\textwidth]{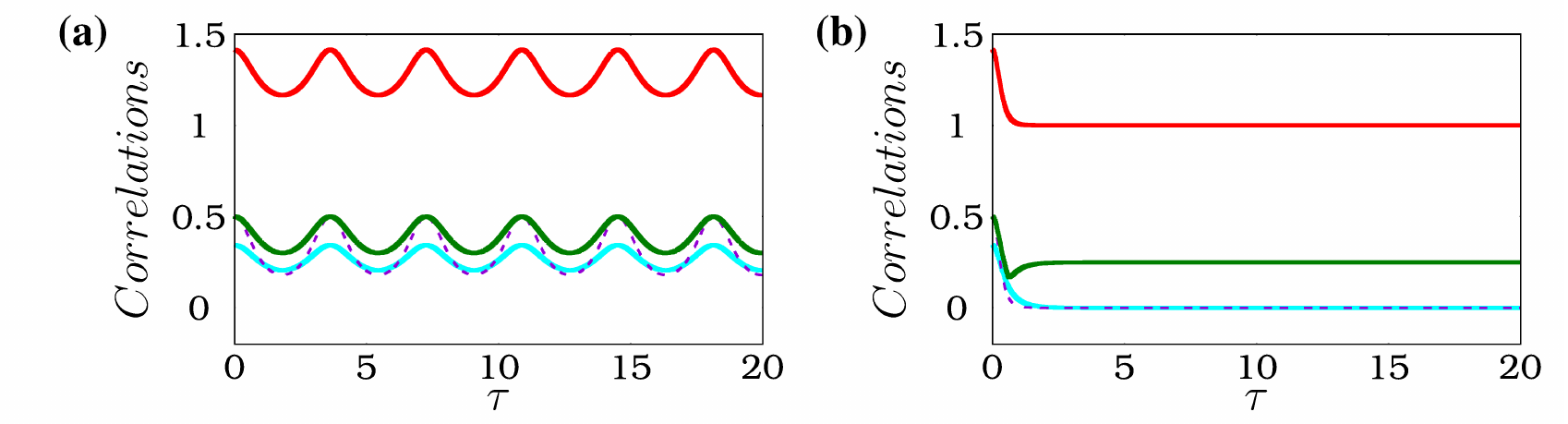}
	\caption{Fig. (a) and (b) shows the comparison of MIN (purple dashed line), Trace MIN (darkgreen solid line), Bell inequalities (Red solid line) and Concurrence (cyan solid line) for two different values of $\tilde{\gamma}=0.5, 1.5$ with $r=0.5$ for a mixed pure state $\ket{\psi (0)}=\frac{1}{\sqrt{2}}(\ket{01}+\ket{10})$.}
	\label{mixed110}
\end{figure*}
\par Fig. 6 shows the comparative dynamics of concurrence ($\mathcal{C}\mathcal{(\rho)}$), MIN ($\mathcal{N}_{2}\mathcal{(\rho)}$), TMIN ($\mathcal{N}_{1}\mathcal{(\rho)}$) and Bell-inequality ($\mathcal{B} (\rho)$) for the mixed state with two different values of $\tilde{\gamma}$. As discussed in the earlier case of mixed state, here also, the correlation measures exhibit periodic oscillations with respect to time.  For the values of $\tilde{\gamma}<1$, the function $\mathcal{C}\mathcal{(\rho)}$, $\mathcal{N}_{2}\mathcal{(\rho)}$ and $\mathcal{N}_{1}\mathcal{(\rho)}$ clearly depict the presence of quantum correlations whereas $\mathcal{B} (\rho)$ does not capture the presence of quantum correlations as reported in Fig. 6(a). Again, the function $\mathcal{B} (\rho)$ is less than 2 for any value of $\tilde{\gamma}$ weakly denied to show the correlations. As seen in Fig. 6(b), the other two functions  $\mathcal{C}\mathcal{(\rho)}$, and $\mathcal{N}_{2}\mathcal{(\rho)}$ turns to be zero indicates the absence of quantum correlations for $\tilde{\gamma}>1$. Nevertheless, the quantum correlation is clearly quantified by the $\mathcal{N}_{1}\mathcal{(\rho)}$ for any given value of $\tilde{\gamma}>1$ as shown in Fig. 6(b). The above observation clearly indicates that the $\mathcal{N}_{1}\mathcal{(\rho)}$ acts as an effective quantifier of bipartite quantum correlation and indicators of nonlocal attributes of a quantum system. 
\section{Conclusions}
\label{cncl}
In conclusion, we have examined the dynamical behaviors of quantum correlations quantified by the entanglement, measurement-induced nonlocality, and Bell nonlocality under local nonunitary evolution. For the pure state, at lower values of $\tilde{\gamma}$, the correlation measures show the periodic oscillation and the higher values of $\tilde{\gamma}$ causes rapid decay of quantum correlation to a steady state value. The trace distance MIN reaches its steady state value after  some finite time while MIN and entanglement decay to zero. Further, the Bell inequality saturates at 2. We observe that the presence of nonlocality in terms of trace MIN,  whereas the Bell inequality and entanglement fail to capture the quantum correlation. For the initially mixed state, we observe that Bell inequality is completely failed to capture the nonlocality even when the qubits are entangled. This also implies a fact that the presence of entanglement does not necessarily indicate the nonlocal attributes of the quantum system. It is observed that the MINs are more robust against nonunitary evolution and pronounced than the Bell inequality. 

Our results also emphasize that the trace distance based measurement-induced nonlocality offers more resistance to nonunitary evolution than the Bell inequality and entanglement. Further, we observe that the time evolution of Hermitian and non-Hermitian systems exhibit similar dynamics.
\section*{Acknowledgements}
JRP thanks the Department of Science and Technology,
Government of India, for providing an INSPIRE Fellowship
No. DST/INSPIRE Fellowship/2017/IF170539. RM acknowledge the financial support from the Council of Scientific and Industrial Research (CSIR), Government of India, under Grant No. 03(1444)/18/EMR-II.  The work of VKC forms part of the research projects sponsored by SERB-DST-MATRICS Grant No. MTR/2018/000676 and CSIR Project Grant No. 03(1444)/18/EMR-II. JRP, RM, and VKC wish to thank DST, New Delhi for computational
facilities under the DST-FIST programme (SR/FST/PS- 1/2020/135) to the Department of Physics.


\begin{thebibliography}{99}
\bibitem{nonhermpt1}
C. M. Bender and Boettcher, Phys. Rev. Lett. {\bf 80}, 5243-5246 (1998).
\bibitem{nonhermpt2}
C. M Bender and Boettcher, and S. Meisinger, J. Math. Phys, {\bf 40}, 2201-2229 (1999). 
\bibitem{nonhermmpt3}
C. M. Bender, D. C. Brody, and H. F. Jones, Phys. rev. Lett. {\bf 89}, 270401 (2002).
\bibitem{nonhemenvir1}
 A. Sergi, and K.G. Zloshchastiev, Int. J. Mod. Phys. B {\bf 27}, 1345053 (2013). 
\bibitem{nonhemenvir2}
 A. Sergi, and K.G. Zloshchastiev, Phys.
Rev. A 91, 062108. 
\bibitem{radiactive}
C. M. Bender, Rep. Prog. Phys. {\bf 70}, 947-1018 (2007).
\bibitem{neutron}
H. Feshbach, C. E. Porter, and V. F. Weisskopf, Phys. Rev. {\bf 96}, 448-464 (1954).

\bibitem{Nielsen2010}
M. Nielsen, and I. Chuang.: Quantum Computation and Quantum Information, Cambridge University Press, Cambridge (2010).

\bibitem{Baumgratz}
T. Baumgratz, M. Cramer, and M.B. Plenio, Phys. Rev. Lett. {\bf 113} 140401 (2014).

\bibitem{Einstein}
A. Einstein, B. Podolsky, and N. Rosen, Phys. Rev. {\bf 47}, 777 (1935).

\bibitem{Schrodinger}
E. Schrodinger, Proc. Cambridge Philos. Soc. {\bf 31}, 555 (1935).

\bibitem{Bell}
J. S. Bell, Physics {\bf 1}, 195 (1964).

\bibitem{Schneeloch}
J. Schneeloch, C.J. Broadbent, S.P. Walborn, E.G. Cavalcanti, and J.C. Howell, Phys.
Rev. A {\bf 87}, 062103 (2013).

\bibitem{Ollivier}
H. Ollivier, and W.H. Zurek, Phys. Rev. Lett. {\bf 88}, 017901 (2001).

\bibitem{Luo2011}
S. Luo, and S. Fu, Phys. Rev. Lett. {\bf 106}, 120401 (2011).

\bibitem{Luo2008}
S. Luo, Phys. Rev. A {\bf 77}, 022301 (2008).

\bibitem{Dakic}
B. V. Daki\'c, and V. Brukner, Phys. Rev. Lett. {\bf 105}, 190502 (2010).

\bibitem{Luo2010}
S. Luo, and S. Fu,  Phys. Rev. A {\bf 82}, 034302 (2010).

\bibitem{Xiong2014}
S. Xiong, W. J. Zhang, C. S. Yu, and H-S. Song   Phys. Lett. A {\bf 378}, 344 (2014).

\bibitem{Jin2012}
J.-S. Jin, F.-Y. Zhang,  C.-S. Yu, and H-S. Song, J. Phys. A: Math. Theor. {\bf 45}, 115308 (2012).

\bibitem{Passante}
G. Passante, O. Moussa, and R. Laflamme, Phys. Rev. A {\bf 85}, 032325 (2012).

\bibitem{Girolami2012}
D.Girolami, and G. Adesso, Phys. Rev. Lett. {\bf 108}, 150403 (2012).

\bibitem{Piani2012}
M. Piani, Phys. Rev. A {\bf 86}, 034101 (2012).

\bibitem{Muthu1}
R. Muthuganesan, and R. Sankaranarayanan   Phys. Lett. A {\bf 381}, 3028 (2017).

\bibitem{Fan2015}
M-L. Hu, and H. Fan, New J. Phys. {\bf 17}, 033004 (2015).

\bibitem{Muthu2}
R. Muthuganesan, and V. K. Chandrasekar,   Commun. Theor. Phys. {\bf 72}, 075103 (2020).

\bibitem{indra2021}
V. S. Indrajith, R. Muthuganesan, and R. Sankaranarayanan, Physica A {\bf 566}, 125615 (2021).

\bibitem{Xi2012}
Z. Xi, X. Wang, and Y. Li, Phys. Rev. A {\bf 85}, 042325 (2012).

\bibitem{Hu2012}
M.-L. Hu, and H. Fu, Ann. Phys. {\bf 327}, 2343 (2012).

\bibitem{Li2016}
L. Li, Q.-W. Wang, S.-Q. Shen, and M. Li, Europhys. Lett. {\bf 114}, 10007 (2016).

\bibitem{indra1}
V.S. Indrajith, R. Muthuganesan, and R. Sankaranarayanan, Physica A {\bf 527}, 121325 (2019).

\bibitem{Mohamed1}
A.-B.A.Mohamed, H.A.Hessian, and H.Eleuch, Chaos, Solitons and Fractals {\bf 135}, 109773 (2020).

\bibitem{Bhuvana}
S. Bhuvaneswari, R. Muthuganesan, and R. Radha, Physica A {\bf 573}, 125932 (2021).

\bibitem{Mohamed2}
A.-B.A. Mohamed, A-H Abdel-Aty, and H. Eleuch,  Physica E {\bf 128},  114529  (2021)

\bibitem{Muthu2021QIP}
R. Muthuganesan, and V. K. Chandrasekar,  Quantum Inf. Process {\bf 20}, 46 (2021).

\bibitem{Wootters}
S. Hill, and W. K. Wootters, Phys. Rev. Lett. {\bf 78}, 5022 (1997).

\bibitem{Brunner}
 N. Brunner, D. Cavalcanti, S. Pironio, V. Scarani, S. Wehner, Rev. Mod. Phys. {\bf 86}, 419 (2014).
 
 \bibitem{Horodecki}
 R. Horodecki, P. Horodecki, and M. Horodecki, Phys. Lett. A {\bf 200}, 340 (1995).



\bibitem{baseref}
S.-Y. Zang, M.-F. Fang, and L. Xu, Quantum Inf Process {\bf 16}, 234 (2017).



\end{thebibliography}
\end{document}